\documentclass[reprint,aps,pra,amsmath,amssymb,superscriptaddress,floatfix]{revtex4-2}

\usepackage{color}
\usepackage{graphicx}
\usepackage{xspace}
\usepackage{dcolumn}
\usepackage{bm}
\usepackage[dvipsnames]{xcolor} 
\usepackage{soul} 
\usepackage{gensymb} 

\newcommand{\equref}[1]{Eq.~(\ref{#1})}
\newcommand{\figref}[1]{Fig.~\ref{#1}}

\definecolor{go_green}{rgb}{0.13, 0.55, 0.13}

\renewcommand{\vec}[1]{\boldsymbol{#1}}
\renewcommand{\approx}{\simeq}
\usepackage{braket}
\newcommand{\pdagger}{{\phantom{\dagger}}}

\usepackage[colorlinks=true]{hyperref}  
\hypersetup{
    bookmarks=true,         
    unicode=false,          
    pdftoolbar=true,        
    pdfmenubar=true,        
    pdffitwindow=false,     
    pdfstartview={FitH},    
    pdftitle={Displacement-field-tunable superconductivity \\ in an inversion-symmetric twisted van der Waals heterostructure},    
    pdfauthor={Scammell and Scheurer},     
    pdfsubject={},   
    pdfcreator={},   
    pdfproducer={}, 
    pdfkeywords={} {} {}, 
    pdfnewwindow=true,      
    colorlinks=true,       
    linkcolor=blue, 
    citecolor=blue,        
    filecolor=magenta,      
    urlcolor=blue           
}

\begin{document}

\title{Anisotropic multiband superconductivity in 2M-WS$_{2}$ probed by controlled disorder}

\author{Sunil Ghimire}
\affiliation{Ames National Laboratory, Ames, Iowa 50011, USA}
\affiliation{Department of Physics \& Astronomy, Iowa State University, Ames, Iowa 50011, USA}

\author{Kamal~R.~Joshi}
\affiliation{Ames National Laboratory, Ames, Iowa 50011, USA}
\affiliation{Department of Physics \& Astronomy, Iowa State University, Ames, Iowa 50011, USA}

\author{Marcin~Ko\'{n}czykowski}
\affiliation{Laboratoire des Solides Irradi\'{e}s, CEA/DRF/lRAMIS, \'{E}cole Polytechnique, CNRS, Institut Polytechnique de Paris, F-91128 Palaiseau, France}

\author{Romain~Grasset}
\affiliation{Laboratoire des Solides Irradi\'{e}s, CEA/DRF/lRAMIS, \'{E}cole Polytechnique, CNRS, Institut Polytechnique de Paris, F-91128 Palaiseau, France}

\author{Amlan~Datta}
\affiliation{Ames National Laboratory, Ames, Iowa 50011, USA}
\affiliation{Department of Physics \& Astronomy, Iowa State University, Ames, Iowa 50011, USA}

\author{Makariy~A.~Tanatar}
\affiliation{Ames National Laboratory, Ames, Iowa 50011, USA}
\affiliation{Department of Physics \& Astronomy, Iowa State University, Ames, Iowa 50011, USA}

\author{Damien B\'erub\'e}
\affiliation{Department of Chemistry and Chemical Biology, Harvard University, Massachusetts 02138, USA}

\author{Su-Yang Xu}
\affiliation{Department of Chemistry and Chemical Biology, Harvard University, Massachusetts 02138, USA}

\author{Yuqiang Fang}
\affiliation{State Key Laboratory of High-Performance Ceramics and Superﬁne Microstructure, Shanghai Institute of Ceramics Chinese Academy of Sciences, Shanghai 200050, P. R. China}

\author{Fuqiang Huang}
\email[Corresponding author: ]{huangfq@pku.edu.cn}
\affiliation{College of Chemistry and Molecular Engineering, Peking University, Beijing, 100871, P. R. China}

\author{Peter P.~Orth}
\affiliation{Ames National Laboratory, Ames, Iowa 50011, USA}
\affiliation{Department of Physics \& Astronomy, Iowa State University, Ames, Iowa 50011, USA}
\affiliation{Department of Physics, Saarland University, 66123 Saarbr\"{u}cken, Germany}

\author{Mathias S.~Scheurer}
\affiliation{Institute for Theoretical Physics, University of Innsbruck, Innsbruck A-6020, Austria}
\affiliation{Institute for Theoretical Physics III, University of Stuttgart, 70550 Stuttgart, Germany}

\author{Ruslan~Prozorov}
\email[Corresponding author: ]{prozorov@ameslab.gov}
\affiliation{Ames National Laboratory, Ames, Iowa 50011, USA}
\affiliation{Department of Physics \& Astronomy, Iowa State University, Ames, Iowa 50011, USA}

\date{\today}

\begin{abstract}
The intrinsically superconducting Dirac semimetal 2M-WS$_{2}$ is a promising candidate to realize proximity-induced topological superconductivity in its protected surface states. A precise characterization of the bulk superconducting state is essential for understanding the nature of surface superconductivity in the system. Here, we perform a detailed experimental study of the temperature and nonmagnetic disorder dependence of the London penetration depth $\lambda$, the upper critical field $H_{c2}$, and the superconducting transition temperature $T_c$ in 2M-WS$_{2}$.
We observe a power-law dependence $\lambda(T) - \lambda(0) \propto T^{3}$ at temperatures below $0.35~T_c$, which is remarkably different from the expected exponential attenuation of a fully gapped isotropic $s$-wave superconductor. We then probe the effect of controlled nonmagnetic disorder induced by 2.5 MeV electron irradiation at various doses and find a significant $T_c$ suppression rate. Together with the observed increase of the slope $dH_{c2}/dT|_{T=T_c}$ with irradiation, our results reveal a strongly anisotropic $s^{++}$ multiband superconducting state that takes the same sign on different Fermi sheets. Our results have direct consequences for the expected proximity-induced superconductivity of the topological surface states.
\end{abstract}
\maketitle

\section{Introduction}
Topological superconductors (TSCs) are characterized by non-trivial topology and superconducting gap opening, which give rise to the emergence of zero energy excitations called Majorana fermions, a particle that is its own antiparticle~\cite{readPairedStatesFermions2000, kitaevUnpairedMajoranaFermions2001, ivanovNonAbelianStatisticsHalfQuantum2001, Nayak_RevModPhys_2008}.  
TSCs are the most promising experimental platform for the realization of Majorana fermion zero modes, which lie at the foundation of topological quantum information science~\cite{kitaevFaulttolerantQuantumComputation2003a, sternTopologicalQuantumComputation2013, Das_npj_2015, lahtinenShortIntroductionTopological2017}. A well-defined theoretical framework of this potential application motivates researchers to extensively search for topological superconductors and Majorana fermions in bulk materials. 

Intrinsic TSCs have topologically nontrivial Bogoliubov bands. This requires a momentum-dependent order parameter, e.g., $p$-wave pairing, and an unconventional pairing mechanism \cite{PhysRevB.93.174509,PhysRevB.90.184512}, making them rare in nature. One of the most studied compounds is Sr$_{2}$RuO$_{4}$  \cite{Kallin_Report_progress_2012, Kallin_JPCM_2009}. However, $p$-wave superconductors are extremely sensitive to disorder, and the realization of topological edge states is still debatable. Other intrinsic TSCs candidates are non-stochiometric compounds such as  Cu$_{x}$Bi$_{2}$Se$_{3}$ \cite{Fu_PRL_2010,Hor_PRL_2010}, Fe$_{x}$Te$_{x}$Se$_{1-x}$ \cite{Xu_PRL_2016,Zhang_Science_2017, Wang_Science_2018},  LiFe$_{x}$Co$_{1-x}$As$_{x}$~\cite{Zhang_NaturePhy_2019} which requires compositional fine tuning. Another way of realizing topological superconductivity is by combining a topological insulator with an $s$-wave superconductor~\cite{fuSuperconductingProximityEffect2008b} or via the proximity effect between a BCS superconductor and a Rashba semiconductor in a magnetic field~\cite{sauGenericNewPlatform2010,aliceaMajoranaFermionsTunable2010, oregHelicalLiquidsMajorana2010, lutchynMajoranaFermionsTopological2010, aliceaNonAbelianStatisticsTopological2011}. In some experiments, Majorana bound states have potentially been realized and interrogated  \cite{Das_NaturePhy_2012,Stevan_Science_2014,Sun_PRL_2016,Albrecht_Nature_2016}, however, these heterostructures require long coherence lengths, making this approach experimentally  challenging.

An alternative approach that avoids sophisticated heterostructure engineering is to consider an intrinsically  superconducting material that exhibits topologically protected surface states in the normal state. Once the bulk material becomes superconducting, it can induce superconductivity in the surface state via the proximity effect. 
Recently, the stochiometric transition metal dichalcogenide (TMD) 2M-WS$_{2}$ has been discovered that becomes superconducting with transition temperature, $T_c \approx 8.8$~K~\cite{Fang_Ad.Mat_2019, Yuan_NatPhy_2019}. High-resolution angle-resolved photoemission spectroscopy (ARPES) and spin-resolved ARPES found gapless surface states with
spin-momentum locking, suggesting a non-trivial topological surface state of this layered Dirac semimetal~\cite{Cho_Nanoletters_2022}. The surface states have been shown to become gapped below $T_c$~\cite{Li_NAtComm_2021}. 
Based on these studies, 2M-WS$_{2}$ is therefore a promising candidate for realizing topological superconductivity on the surface. As for the bulk superconducting properties, a $\mu$SR study concluded that 2M-WS$_2$ exhibits nodeless superconductivity. Thermal conductivity measurements also support a nodeless superconducting gap in 2M-WS$_{2}$ \cite{Wang_PRB_2020}. Finally, a theoretical study using \textit{ab-initio} Migdal-Eliashberg theory suggested an anisotropic but full-gap superconducting order parameter with $s$-wave symmetry for both bulk and bilayer 2M-WS$_{2}$ \cite{Lian_NanoLetters_2020}. 

Since the symmetry and detailed variation of the bulk superconducting order parameter on the Fermi surface are expected to have important impacts on the possible proximity-induced superconductivity of the topological surface states, we here provide a detailed experimental study of the superconducting pairing state in single crystals of 2M-WS$_{2}$. We report low-temperature measurements  of the London penetration depth $\lambda(T)$, which is sensitive to thermally-excited low-energy quasiparticles. We also probe the nature of the superconducting state by introducing non-magnetic disorder, in a controlled way, by 2.5 MeV electron irradiation. Our results show $\Delta \lambda(T) = \lambda(T) - \lambda(0) \sim T^3$ at low temperatures in pristine samples. We subsequently irradiated the sample with 2.5 MeV electrons of different doses up to a maximum accumulated dose of $3 \textrm{C}/\textrm{cm}^2=1.87 \times 10^{19}\:\textrm{e}^{-}/\textrm{cm}^{2}$. The temperature dependence of $\Delta \lambda(T)$ remains close t  o $T^3$. Surprisingly, 
we observed a significant rate of $T_c$ suppression from 8.5~K to 5.2~K at the maximum accumulated dose. 
Having different intermediate doses, we explore $T_c$ versus scattering rate, $\Gamma$ (defined in detail below), comparing it with theory and previous studies. Our results provide strong evidence for a highly anisotropic yet still fully gapped superconducting order parameter. 
We also measure the dependence of $H_{c2}(T)$ before and after maximal irradiation and find that the slope $d H_{c2}{dT}|_{T=T_c}$ \emph{increases} after irradiation. As we show in detail below, these two observations together strongly suggest an anisotropic $s^{++}$ pairing state (rather than an $s^{+-}$ state) that exhibits the same sign on different Fermi sheets. 

\section{Experimental}
 
Single crystals of 2M-WS$_{2}$ were prepared by deintercalation of interlayer potassium cations from K$_{0.7}$WS$_{2}$ crystals \cite{Fang_Ad.Mat_2019}. To obtain K$_{0.7}$WS$_{2}$, K$_{2}$S$_{2}$ (prepared using liquid ammonia), W (99.9\% from Alfa Aesar) and S (99.99\% from Alfa Aesar) were mixed in required stoichiometric
proportions and ground in an argon-filled glovebox. The mixtures were pressed
into a pellet and sealed in the evacuated quartz tube. The tube was heated
at 850 \celsius\xspace for 2000 min and slowly cooled to 550 \celsius\xspace at a rate of 0.1 \celsius/min. The synthesized K$_{0.7}$WS$_{2}$ (0.1 g) was oxidized chemically by K$_2$Cr$_2$O$_7$ (0.01 mol L$^{-1}$) in aqueous H$_2$SO$_4$ (50 mL, 0.02 mol/L) at room
temperature for an hour. Finally, the 2M-WS$_{2}$ crystals were extracted after
washing in distilled water several times and drying in a vacuum oven at room temperature.

The London penetration depth was measured using a sensitive tunnel-diode resonator (TDR) operated at a radio frequency of around 10 MHz. The technique is described in detail elsewhere \cite{VanDegrift1975RSI,Prozorov2000PRB,Prozorov2000a,Prozorov2006,Prozorov2021}. The sample is subject to a small, $H_{ac}< 20~\text{mOe}$, ac magnetic field. The resonant frequency changes are proportional to the sample's magnetic susceptibility, thus, with proper calibration, to the magnetic penetration depth. The calibration for realistic non-ellipsoidal samples is described in Ref.~\cite{Prozorov2021}. The base temperature achieved in this study was 400~mK ($\sim 0.05~T_c$ of the pristine sample).

The 2.5 MeV electron irradiation was performed at the ``SIRIUS" electrostatic accelerator facility at Laboratoire des Solides Irradi\'{e}s, \'{E}cole Polytechnique, Palaiseau, France. At the energy of 2.5 MeV, electrons are moving with the relativistic speed of $0.985c$ with a total current of about 2.7 $\mu$A through a 5 mm diameter diaphragm. The acquired irradiation dose is measured by a calibrated Faraday trap behind the sample and is conveniently expressed in $\textrm{C}/\textrm{cm}^2$, where 1 $\textrm{C}/\textrm{cm}^2=6.24\times10^{18}\:\textrm{e}^{-}/\textrm{cm}^2$. This dose is acquired overnight. Electrons are particularly useful among other particles used for irradiation to induce defects. Unlike heavier particles (heavy ions, protons, $\alpha-$particles) which result in columnar tracks of extended dendritic cascades, electrons produce well-separated point-like defects called Frenkel pairs (vacancy+interstitial) because their relativistic energy transfer upon collisions matches the threshold knock out energy, typically between 20 and 80 eV. On the other hand, thanks to their charge and velocity, their scattering cross-section is reasonable, of the order of 100 barns, compared to neutrons and gamma-rays, also used for this purpose. Another important parameter is the penetration depth into the material. While heavier particles require very thin samples, electrons can be used for typical single crystals. The produced defects are in the dilute limit, typically with about one defect per thousand lattice ions. Except for special circumstances (e.g., semiconductors), they do not ``charge dope" the system and do not cause a shift of the chemical potential. Careful Hall effect measurements verified this. The science of electron damage in materials is well developed \cite{Damask1963,Thompson1969}.

The irradiation has to be conducted at low temperatures to remove heat generated upon collisions and to prevent recombination and, importantly, clusterization of resultant defects. Interstitials have much lower diffusion energy than vacancies, so they migrate to various sinks, such as lattice defects, dislocations, and surfaces, leaving a metastable population of vacancies behind. In our case, the irradiated sample is immersed in liquid hydrogen at about 22 K. Upon warming up, some defects recombine at a material-dependent rate. Therefore the indicated doses are only used to compare irradiations for the same type of samples with the same annealing rate. In-situ resistivity measurements at 22 K showed that resistivity is linearly proportional to the dose. The actual induced damage is estimated from the changes of residual resistivity to extract dimensionless scattering rate, $\Gamma =\hbar /\left( 2\pi k_BT_{c0}\tau \right)$ \cite{Tanatar2022a}. Here $T_{c0}$ is the superconducting transition temperature in the clean limit and $\tau$ is the transport scattering time. Importantly, the irradiation was carried out on the same samples measured before the irradiation.

\begin{figure}[tb]
\centering
\includegraphics[width=0.9\linewidth]{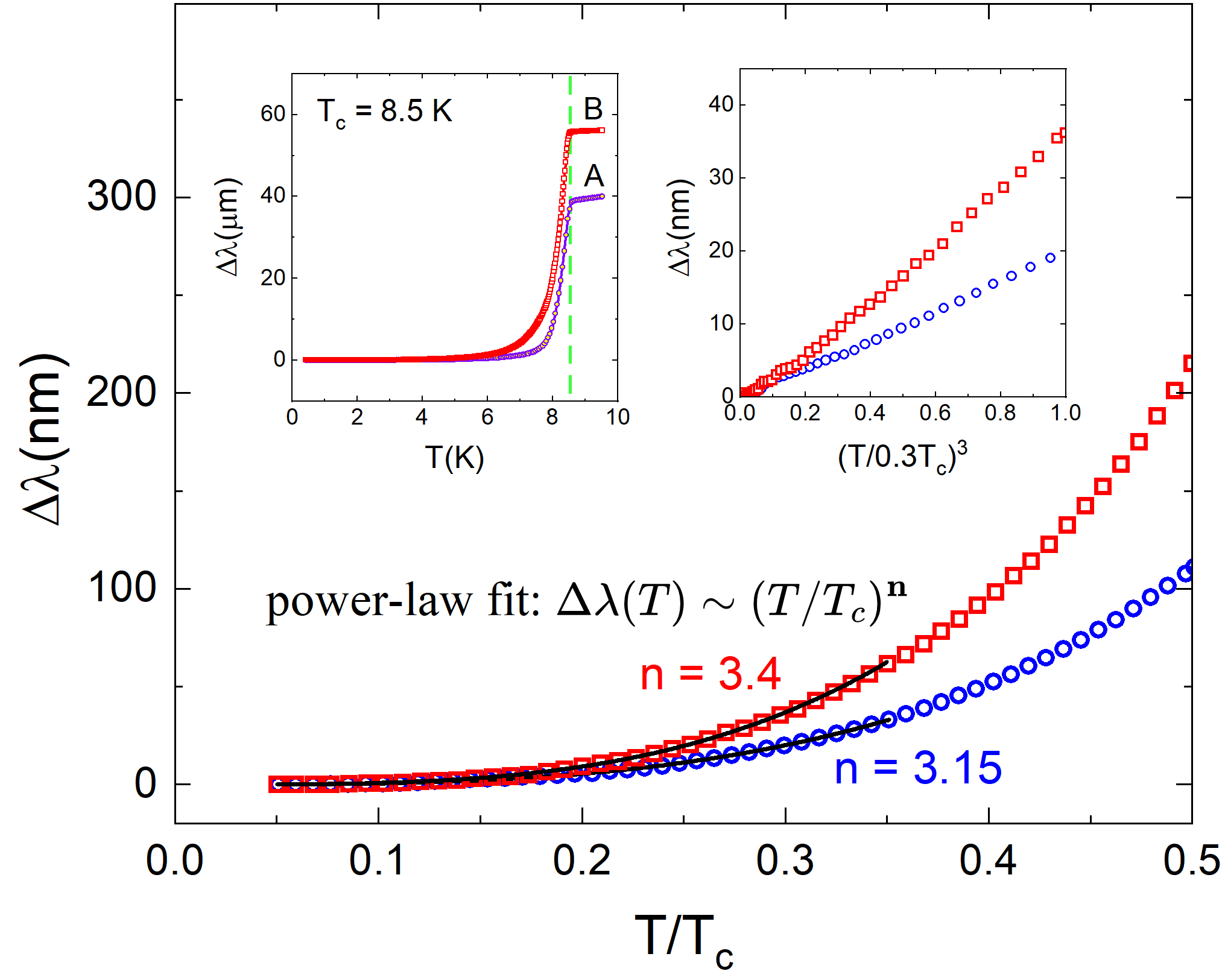}
\caption{Temperature dependence of the London penetration depth, plotted as a departure from its value at the base temperature of 0.4 K. Two pristine samples, (A) and (B), are shown. The main panel focuses on the low-temperature region. Black solid lines show the best fit in the interval $T_{\textrm{min}}\leq T \leq 0.35~T_c$, using the power-law, $\Delta \lambda (T)=A+BT^n$, with $n=3.15$ obtained for sample A, and $n=3.4$ for sample B. The top left inset presents data in the full temperature interval showing sharp and uniform superconducting transition in both samples with $T_c \approx 8.5$~K. The top right inset shows data in the fitting range,  $T \leq 0.35T_c$, plotted as a function of $(T/0.3T_c)^3$ and demonstrating a practically perfect cubic variation of $\lambda(T)$.}
\label{fig1}
\end{figure}

\section{Results}
\noindent
\textbf{Penetration depth.} Figure~\ref{fig1} shows the temperature dependence of the change of the London penetration depth in pristine samples A (blue circles) and B (red squares), in reduced coordinates, $T/T_c$. The main frame focuses on the low-temperature part showing a power-law fit, $\Delta \lambda (T)=A+BT^n$, in the temperature range from 0.4 K to 3.0 K, where the upper limit corresponds to $0.35T_c$. Fitting results in the exponents $n=3.15$ for sample A, and $n=3.4$ for sample B. This means the energy gap does not have symmetry-imposed nodes because a point node would give $n=2$, and a line node would give $n=1$. The presence of disorder cannot explain such large exponents, recalling that the nodal dirty limit corresponds to $n\leq 2$. There is another way to demonstrate the power-law behavior, shown in the right inset in Fig.~\ref{fig1}. Here $\Delta \lambda$ is plotted versus dimensionless $(T/0.3T_c)^3$ revealing a straight line. Of course, this procedure ``compresses" low temperatures where we do have deviations from pure $T^3$ behavior but this does not affect the overall behavior.

An immediate reaction to the observed $n=3$ would be that this is precisely what has been predicted for topological superconductors \cite{Wu_PRL_2020,Wu_PRB_2021}. However, according to our current understanding \cite{Li_NAtComm_2021,Yuan_NatPhy_2019}, WS$_2$ is not a three-dimensional topological superconductor but instead a topological semi-metal with surface states; when entering the superconducting phase, also the surface states become superconducting, with roughly the same gap \cite{Li_NAtComm_2021}, and can realize a two-dimensional topological superconductor. At least in the clean limit, this surface system has no edges and, thus, no topological zero-modes; the latter will only appear in the presence of defects, such as magnetic vortices \cite{Yuan_NatPhy_2019}. Therefore, the mechanism of \cite{Wu_PRL_2020,Wu_PRB_2021} yielding a $T^3$ contribution does not apply unless (i) there is a high density of magnetic vortices or other defects trapping Majorana modes and (ii) WS$_2$ is indeed a three-dimensional topological superconductor. While (i) seems very unlikely, given the small magnetic fields in our experiments ($\approx 20~\text{mOe}$), (ii) would be extremely exotic and not in line with previous experiments. Therefore, we interpret our findings as consequences of the bulk superconducting phase and not in terms of gapless low-energy excitations at the surface. 

Since WS$_2$ is not easy to obtain in the bulk form, we put special attention to the sample quality screening many samples and finally selecting two with the sharpest transitions shown in this work. The left inset in Fig.~\ref{fig1} shows $\Delta \lambda (T)$ for both samples in the full temperature range consistent with the high quality of the studied samples. The curves do not show any extra features, and the transitions are well defined with the onset for both samples indicating the superconducting transition at $T=8.5$\:K. 

\begin{figure}[tb]
\centering
\includegraphics[width=0.9\linewidth]{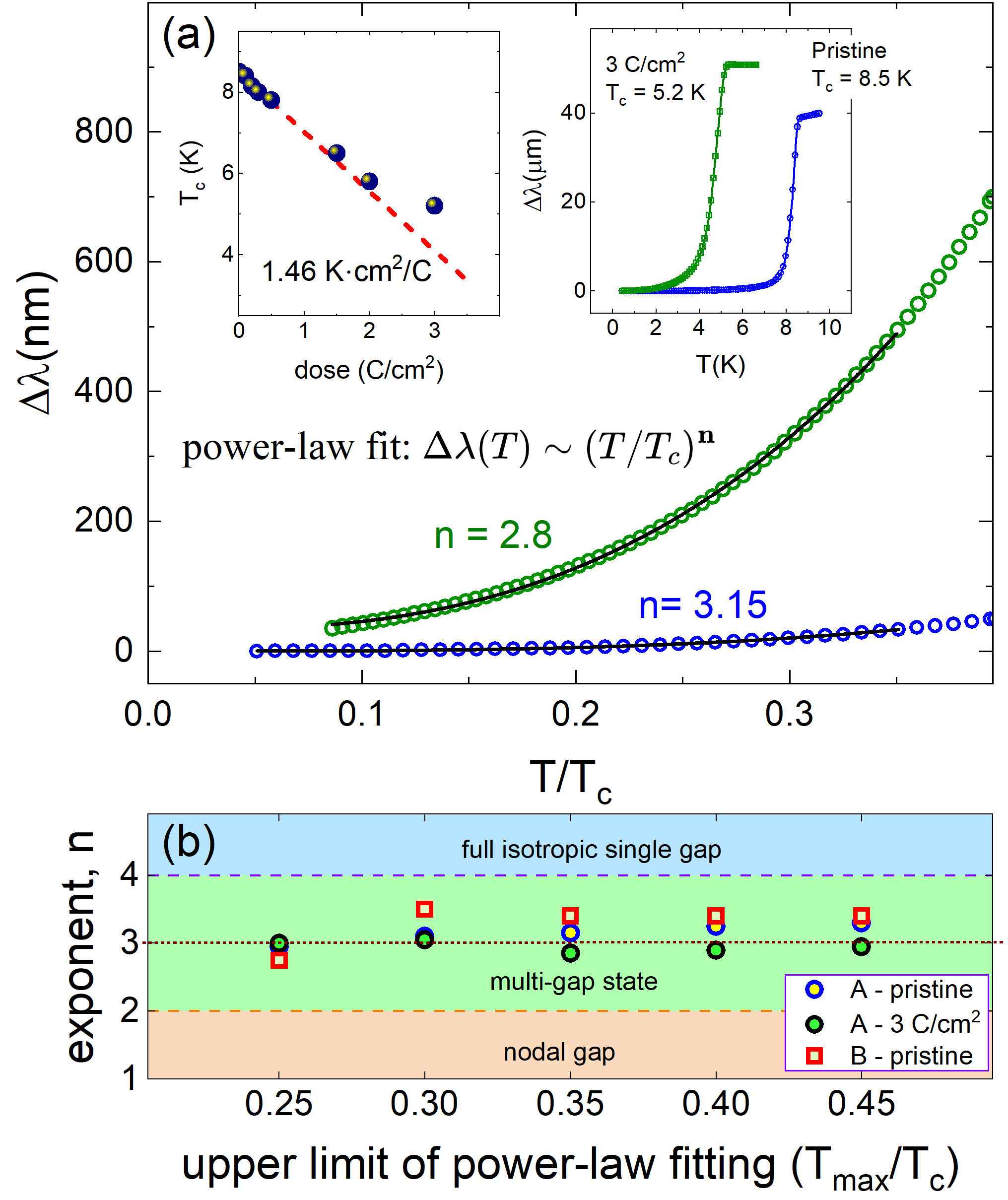}
\caption{(a) $\Delta \lambda (T)$ as a function of normalized temperature $T/T_c$ for sample A before (blue circles) and after (green squares) electron irradiation with the total accumulated dose of  $3 \textrm{C}/\textrm{cm}^2=1.87\times10^{19}\:\textrm{e}^{-}/\textrm{cm}^2$. Black solid lines show the best power-law fit (below  $0.35T_c$), with exponent $n=3.15$ and $n=2.8$, before and after electron irradiation, respectively. Top left inset in panel (a) shows a progression of $T_c$ suppression with the increasing accumulated dose obtained from \textit{in-situ} resistivity measurements. Top right inset displays data over the whole temperature range showing the change in the $\Delta \lambda (T)$ curve. Besides dramatic shift, note that the transition did not broaden and also the value above $T_c$ increased after irradiation due to an increase of the normal-state skin depth ($T>T_c$). Panel (b) shows the exponent $n$ obtained from the power-law fit with a variable upper fit limit. Practically constant value mean the robust $T^3$ behavior of $\lambda(T)$ in the wide region.}
\label{fig2}
\end{figure}

\vspace{1em}
\noindent
\textbf{Controlled disorder.} Next, we turn to the effects of controlled disorder induced by 2.5 MeV electron irradiation. Due to a limited beam time we chose to irradiate only sample A, which showed the best superconducting transition; see the left inset in Fig.~\ref{fig1}. The sample was irradiated in liquid hydrogen at 22 K, extracted to room temperature, and its resistivity was measured ex-situ as a function of temperature using the standard four-probe method. Importantly, the sample was mounted on a Kyocera chip, so the contacts were not disturbed in this procedure. This cycle was repeated a number of times as shown in Fig.~\ref{fig2}. Top panel (a) in Fig.~\ref{fig2} shows the low-temperature variation of the London penetration depth change, $\Delta \lambda (T)$ in sample A plotted vs. reduced temperature, $T/T_c$. Results for the pristine state are shown by blue circles; measurements after the largest accumulated dose of $3 \textrm{C}/\textrm{cm}^2=1.87\times10^{19}\:\textrm{e}^{-}/\textrm{cm}^2$ are shown by green squares. For illustration, the irradiated state curve was shifted vertically, so it extrapolates to zero at $T\rightarrow0$, which does not change the power-law exponent. 

The first striking observation is that the superconducting transition temperature is shifted by 3.3 K from 8.5 to 5.2 K, which is almost 40\% of the pristine-state value. This is remarkable and our second direct indication of the unusual nature of the bulk superconducting order parameter since, to the best of our knowledge, such a rate of $T_c$ suppression has never been reported in any fully gapped superconductor with a single order parameter.
The upper left inset in panel (a) in Fig.~\ref{fig2} shows the summary of the $T_c$ suppression as the function of irradiation dose. It decreases with the initial slope of about $1.46 K\cdot \textrm{cm}^2/\textrm{C}$. The upper right inset shows the full temperature range variation of $\Delta \lambda (T)$. An increase of the apparent penetration depth above $T > T_c$ upon irradiation is due to an expected increase of the normal-state skin depth resulting from the increased residual resistivity. Of course, with decreased $T_c$, the $0.3T_c$ threshold for low-temperature asymptotic behavior decreases in absolute temperature units. Nevertheless, we still obtain the large exponent, $n=2.8$. Note that the transition to the normal state remains sharp after the irradiation.  
This indicates a uniform distribution of point-like scattering centers as we anticipated by estimating the electrons' penetration profiles.

To examine how robust the power-law fit is, panel (b) in Fig.~\ref{fig2} shows the exponent $n$ plotted versus the upper limit of the power-law fitting, $T_{\text{max}}/T_c$. The lower limit was the lowest possible, $T_{\text{min}}/T_c$, which is equal to $0.047$ in pristine samples, and to $0.077$ in the irradiated with the dose of $3 \textrm{C}/\textrm{cm}^2$. The result shows that the exponent is quite robust and, statistically, does not change between these samples clustering around $n=3$. For illustration, the regime of a fully gapped isotropic state is shown at around $n\geq 4$, which is numerically indistinguishable from the exponential attenuation. On the other side, the nodal scenarios, dirty or clean, point or line nodes, are located below $n=2$. We, therefore, have quite an unusual behavior, - robust against the disorder functional temperature dependence of the London penetration depth $\sim T^3$ but significant suppression of $T_c$. 

\begin{figure}[tb]
\centering
\includegraphics[width=0.9\linewidth]{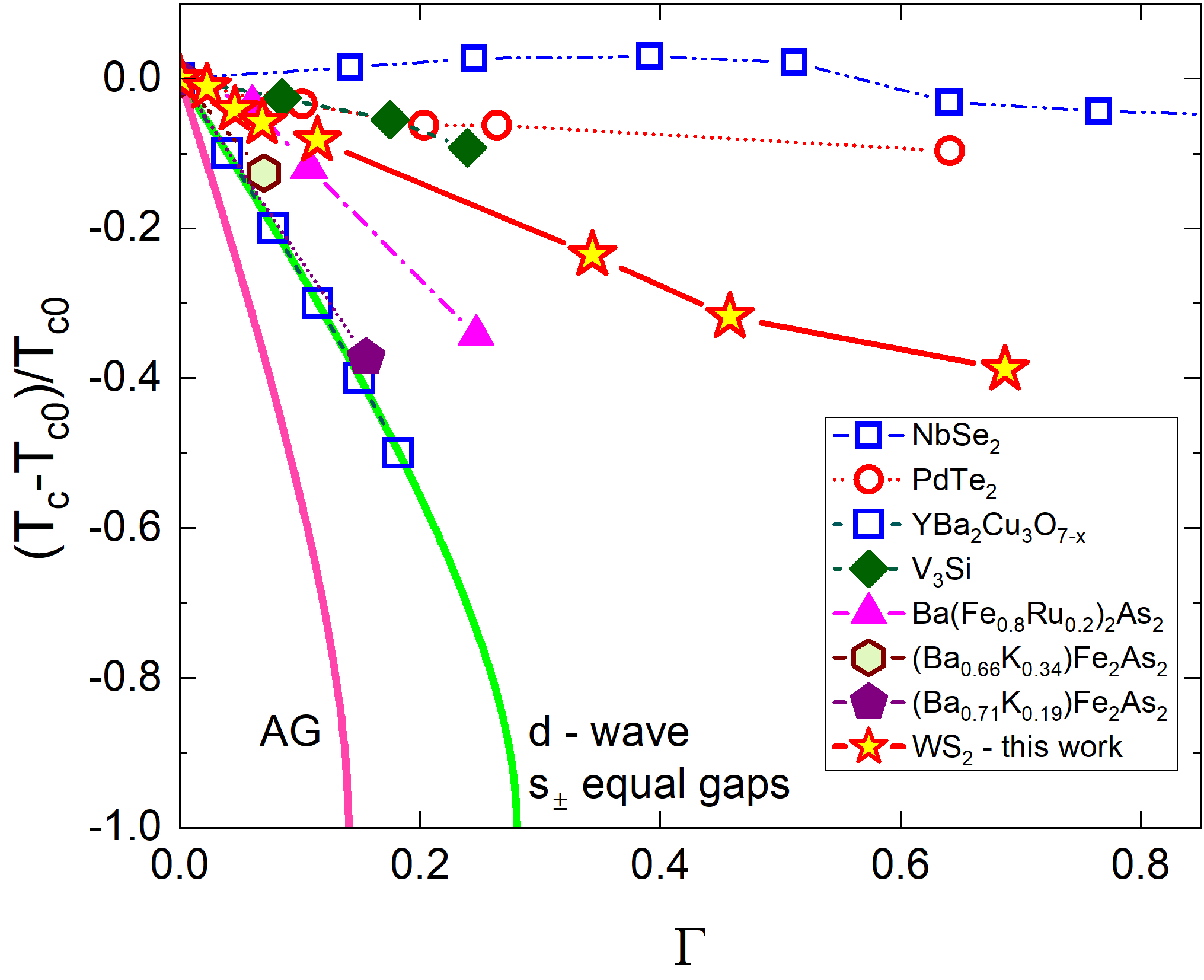}
\caption{Normalized change of the superconducting transition temperature, $\Delta t_{c} = (T_c-T_{c0})/T_{c0}$ upon electron irradiation in 2M-WS$_{2}$ crystal compared to other superconductors. Dimensionless scattering rate, $\Gamma$ is calculated from the increase of resistivity upon irradiation, Eq.~\eqref{gamma}. The pink curve labelled "AG" refers to the Abrikosov-Gor'kov result~\cite{AbrikosovGorkov1960} for $T_c$ suppression of an isotropic $s$-wave state with increasing magnetic disorder. }
\label{fig3}
\end{figure}

\vspace{1em}
\noindent
\textbf{Dimensionless scattering rate.} To compare the rate of $T_c$ suppression in different materials, with very different initial transition temperatures, $T_{c0}$, we plot in Fig.~\ref{fig3} the normalized $\Delta t_{c} = (T_c-T_{c0})/T_{c0}$ versus the dimensionless scattering rate, $\Gamma$, determined from the increase of resistivity, $\Delta \rho$, after the irradiation. Using Drude and London models, we estimate the scattering time, $\tau^{-1}=\rho\mu_{0}^{-1}\lambda_{0}^{-2}$, where $\mu_{0}=4\pi\times10^{-7}\:$H/m is the magnetic permeability of free space and $\rho$ is resistivity. Therefore, the scattering rate is
\begin{multline}
  \Gamma=\frac{\hbar}{2\pi k_{B}T_{c0}\tau}=\frac{\hbar}{2\pi\mu_{0}k_{B}}\frac{\rho}{T_{c0}\lambda_{0}^{2}}\approx \\ \approx 9673.9\frac{\rho\left[\mathrm{\mu\Omega\cdot cm}\right]}{T_{c0}\left[\mathrm{K}\right]\lambda_{0}^{2}\left[\mathrm{nm^{2}}\right]}.
    \label{gamma}
\end{multline}
With the parameters of our sample A, we estimate $\Gamma=0.23$ per 1 C/cm$^{2}$. The suppression of $T_{c}$ by non-magnetic scattering is not expected for isotropic $s$-wave pairing (Anderson theorem \cite{Anderson1959, abrikosov1959superconducting, abrikosov1959theory}); but it occurs in anisotropic materials~\cite{markowitzEffectImpuritiesCritical1963, hohenberg1964anisotropic, golubovEffectMagneticNonmagnetic1997}. Usually, the single-gap anisotropies are not large, and a significant suppression is not expected. However, in multiband materials, two different gaps can formally be treated as an anisotropic single-gap case on a generalized Fermi surface spanning all bands \cite{golubovEffectMagneticNonmagnetic1997, Timmons2020, Cho2022a}. In that case, any suppression rate is possible from none in an isotropic $s^{++}$ with two equal gaps to a complete suppression at the finite critical value of the scattering rate, $\Gamma=0.28$, in case of $s^{+-}$ with two equal gaps of opposite sign. This is the same as in a $d$-wave superconductor. Iron-based superconductors are a good example of $s^{+-}$ with two effective gaps with different ratios of their amplitudes showing various suppression rates \cite{Cho2018}. 

Examining Fig.~\ref{fig3}, we find that the rate of suppression of the $T_c$ of WS$_2$ is stronger than in the $s^{++}$ cases but is quite similar to some of the iron pnictide superconductors. More quantitatively, we extract a dimensionless disorder sensitivity parameter \cite{Timmons2020} $\zeta \approx 0.18$ for $\Gamma\rightarrow 0$. We also find the tendency to saturation at large $\Gamma$, which is a feature of superconductors with unequal gaps. In a simple picture, non-magnetic scattering averages the superconducting gap, and if the mean value of the total gap is not zero, it determines the dirty limit $T_c$ \cite{Cho2022a}.

\begin{figure}[tb]
\centering
\includegraphics[width=0.9\linewidth]{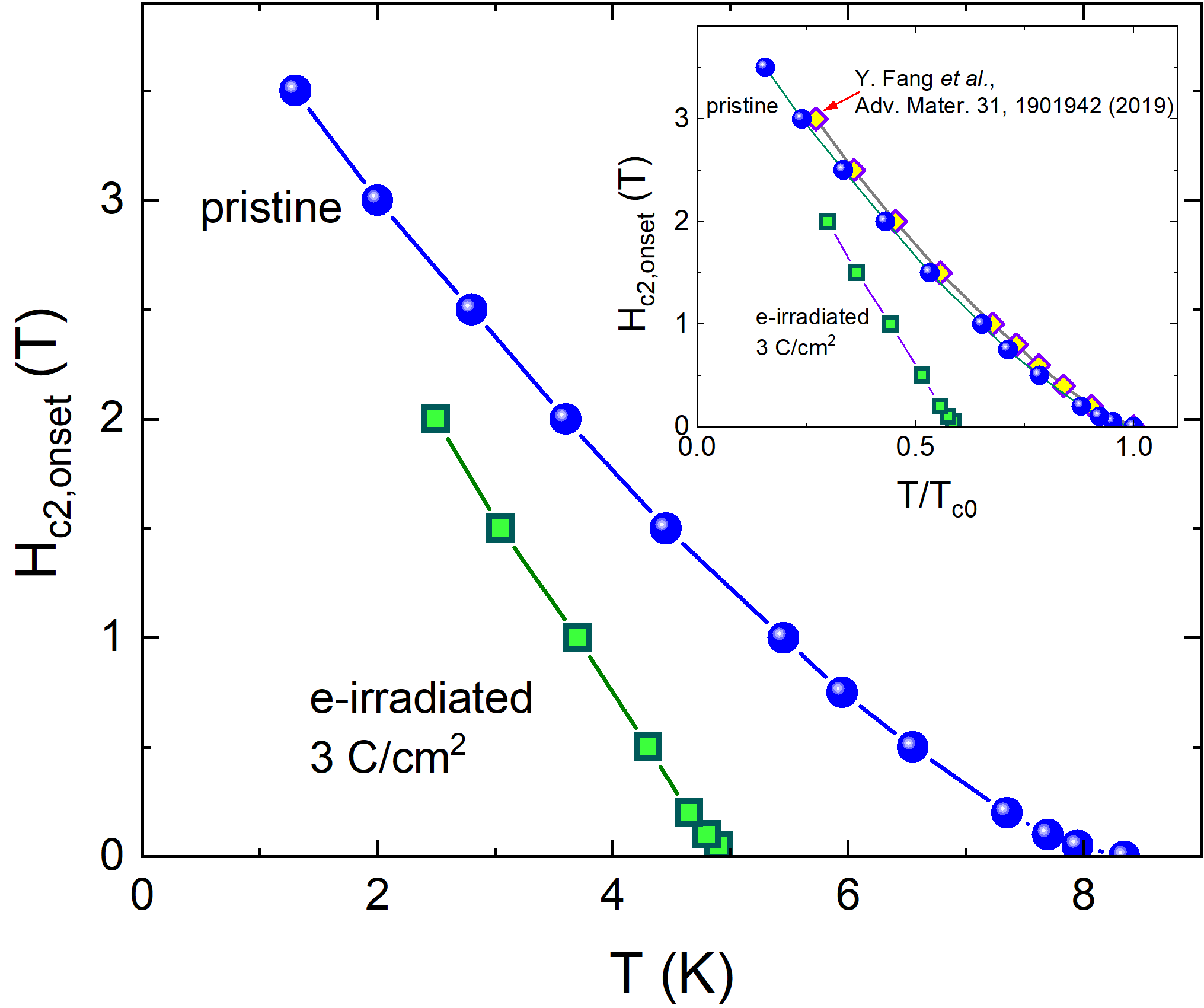}
\caption{Upper critical field, $H_{c2}(T)$, as a function of temperature, measured in sample A before (blue circles) and after (green squares) electron irradiation with the dose of 3 C/cm$^2$. Inset shows the same data but plotted against the normalized temperature, $T/T_{c0}$, by the transition in the pristine state. For comparison, $H_{c2}$ from the discovery paper, Ref.~\cite{Fang_Ad.Mat_2019} showing excellent agreement with our data.}
\label{fig4}
\end{figure}

\vspace{1em}
\noindent
\textbf{Critical field.} With this interesting observation, we now examine another quantity sensitive to disorder scattering---the upper critical field, $H_{c2}$. Figure~\ref{fig4} shows $H_{c2}(T)$ as a function of temperature, measured in sample A before (blue circles) and after (green squares) electron irradiation with the total accumulated dose of 3 C/cm$^2$. The upper critical field was detected as the onset of a TDR frequency shift measured as a function of temperature in different applied DC magnetic fields. The inset in Fig.~\ref{fig4} shows the same data but plotted against the normalized temperature, $T/T_{c0}$, where $T_{c0}$ is the superconducting transition in the pristine state. For comparison, $H_{c2}$ from the WS$_{2}$ discovery paper \cite{Fang_Ad.Mat_2019} is plotted in the same inset showing excellent agreement with our data.

Since, in many cases, it is difficult to measure the entire curve, $H_{c2}\left(T\right)$, and in some cases, the low-temperature limit could be complicated by Pauli limiting physics \cite{Fulde1973} or Fulde-Ferell-Larkin-Ovchinnikov (FFLO) state  \cite{FuldeFerrell1964,LarkinOvchinnikov1964}, since the 1960s the researchers analyzed the slope $dH_{c2}/dT$, at $T_{c}$ \cite{HW1964,HW1966,WHH1966} where the critical field is always orbital (vortices) limited. In addition to the decreased $T_{c}$, already discussed above, Fig.~\ref{fig4} shows that the slope $dH_{c2}/dT$, at $T_{c}$ becomes substantially larger after the irradiation. A standard approach would be to use the HW and claim that this is expected and that the zero-temperature value of $H_{c2}(0)$ has increased as expected from non-magnetic scattering \cite{Kogan2022cf}. However, HW analysis is only valid for isotropic Fermi surfaces and  $s$-wave order parameters. In the case of anisotropies, the situation is significantly more complicated. It has been recently addressed in Ref.~\cite{Kogan2023slope} where it was shown that the slope $dH_{c2}/dT$, at $T_{c}$ increases in case of nodeless superconductors and decreases if there are nodes in the gap function. Ref.~\cite{Prozorov2023slope} has extended this approach to multiband materials and showed that in the case of $s^{+-}$ pairing, this slope should decrease with non-magnetic scattering. Note that in the clean limit, the slope is proportional to $T_c$. However, we are not in the clean limit in the irradiated sample where $T_c$ decreases. The observed reduction shown in Fig.~\ref{fig3}, seem to saturate with increasing scattering rate. This behavior is expected for sufficiently anisotropic single gap or gaps of different magnitudes in the multi-gap scenario. Examining Fig.~\ref{fig4} we see that the slope $dH_{c2}/dT|_{T\rightarrow T_c}$  increased upon irradiation. A similar trend is found in two established multigap $s^{++}$ superconductors, V$_3$Si and NbSe$_2$. This is consistent with the $s^{++}$-pairing, but not with $s^{+-}$. 

\section{Discussion} 

\begin{figure*}[tb!]
\centering
\includegraphics[width=\linewidth]{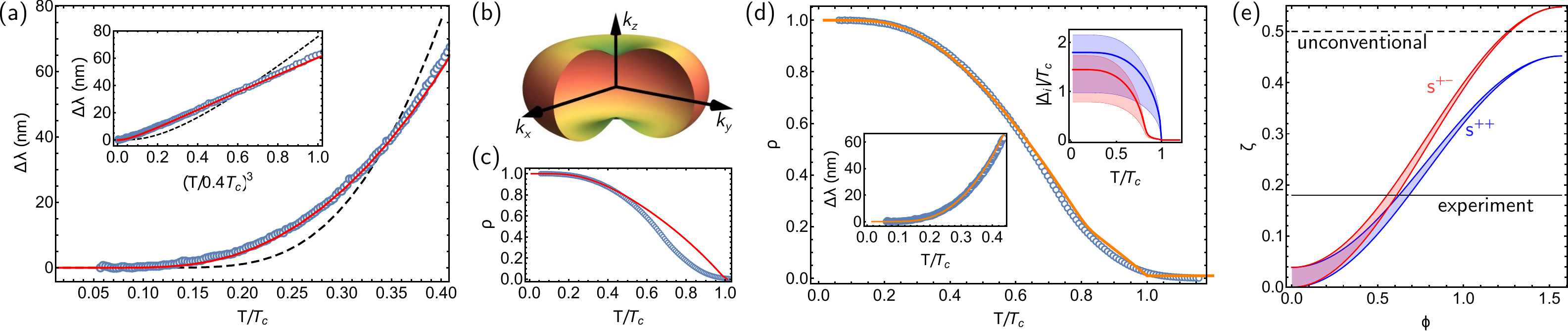}
\caption{\textbf{Theoretical analysis of experimental data.} Part (a) shows the comparison of the anisotropic single-gap (red line, $a=0.49$) and isotropic (black dashed) model with the measured $\Delta \lambda$ in the low-$T$ regime, using a linear (cubic) $T$ axis in the main panel (inset). The anisotropic gap corresponding to $a=0.49$ is illustrated in (b). While the low-temperature behavior is reproduced well, the measured form of $\rho$ for $0.6< T/T_c < 1$ differs significantly from the model prediction, see (c).  This is different for the anisotropic ($a=0.38$) two-band model, see main panel and lower left inset in (d), with temperature dependent gap variations shown in the upper right inset of (d). In (e), we show the spread [for varying $\vartheta$ in \equref{FormScatteringMatrix}] of the dimensionless disorder sensitivity parameter for non-sign-changing ($s^{++}$) and sign-changing ($s^{+-}$) superconducting order parameters; the experimental value is shown as solid line and the black dashed line refers to the case of a superconducting order parameter transforming under a non-trivial representation and impurity scattering between all states on the Fermi surface equally. For concreteness, we here focus on data from sample B. The comparison is similar for sample A.}
\label{theoryfigure}
\end{figure*}

For a more quantitative and systematic discussion of our data, we return to the low-temperature behavior of the penetration depth $\lambda(T)$ of the pristine sample. As can be seen in \figref{theoryfigure}(a), an isotropic single-gap (BCS) superconductor (black dashed line), i.e., with momentum independent order-parameter magnitude $|\Delta_{\vec{k}}| = \Delta_0$, cannot describe the data. As mentioned above, our measurements are incompatible with the exponential temperature dependence of $\Delta \lambda(T)$ in such an isotropic state. However, even in conventional superconductors, $|\Delta_{\vec{k}}|$ generically depends on momentum due to the presence of a lattice. Taking the form, $|\Delta_{\vec{k}}| = \Delta_0 (1 - a \cos 2 \theta_{\vec{k}})$, where $\theta_{\vec{k}}$ is the polar angle of $\vec{k}$---consistent with pairing in the trivial representation of the point group $C_{2h}$ of 2M-WS$_2$---we find good agreement [red solid line in \figref{theoryfigure}(a)] with the measured penetration depth for $a=0.49$. This corresponds to a smaller (but still finite) gap along the $k_z$ direction compared to the ``equator'' in the $k_{x}$-$k_y$ plane, illustrated in \figref{theoryfigure}(b). The minimal gap value is approximately $34\%$ of the maximum value.

While the behavior at low temperatures is well reproduced by the anisotropic single-gap model, this is not the case in the regime with temperatures between $T_c/2$ and $T_c$, where the temperature dependence of the gap magnitude is non-negligible. Using the normalized superfluid density $\rho = (\lambda(0)/\lambda(T))^2$ to quantify the quality of the fit, we observe clear deviations in this temperature regime in \figref{theoryfigure}(c). The data thus implies significant deviations of the temperature variation of the gap from those of single-gap superconductors. To capture this behavior theoretically, we consider a two-band ($j=1,2$) model with intra-band Cooper channel interactions $U_j$ and inter-band interaction $J$ (see Methods). We solve for the temperature dependence of the two gap magnitudes self-consistently, while keeping the momentum dependence of the order parameter on each of the two Fermi surfaces to be of the form $\Delta_{j,\vec{k}} = \Delta_j (1 - a \cos 2 \theta_{\vec{k}})$ for simplicity. We find good agreement with experiment when one of $U_{1,2}$ is slightly larger than the other while both being much larger than $J$ (explicitly $U_1/U_2 \approx 1.098$, $|J|/U_1 \approx 0.0049$), see \figref{theoryfigure}(d); in this regime the smaller gap, associated with the lower $U_j$, exhibits significant deviations in its temperature variation from the standard square-root behavior close to $T_c$ [cf.~upper right inset in \figref{theoryfigure}(d)]. This results from the weak admixing of the smaller gap in the regime between the actual $T_c$ and the hypothetical $T_c$ of the band with the weaker interaction in the uncoupled limit $J\rightarrow 0$. These deviations can capture the behavior of the data close to $T_c$, while leaving the low-temperature behavior mostly unaffected and, thus, describe the data well in the entire temperature range. We extract a slightly smaller asymmetry parameter $a=0.38$ such that the largest gap on a given Fermi surface is about twice large as the smallest gap value and a zero-temperature penetration depth value of $\lambda(0) \approx 644\,\text{nm}$.

To further check this picture, we next compare the theoretical predictions with results of our irradiation experiments. First, we focus on weak disorder ($\Gamma \rightarrow 0$ in \figref{fig3}) and investigate the theoretical expectations for the initial slope of the critical temperature with disorder, as encoded in the dimensionless disorder sensitivity parameter \cite{Timmons2020} $\zeta \approx 0.18$. Focusing on time-reversal and spin-rotation invariant disorder configurations and point-like disorder with momentum-independent matrix elements in each band, impurity scattering can be described by a real and Hermitian $2 \times 2$ matrix $w$ in band space; its diagonal  components describe intra-band scattering amplitudes and its off-diagonal ones describe inter-band scattering. We parametrize the matrix using two angles $\vartheta$ and $\phi$ according to
\begin{equation}
    w = w_0\begin{pmatrix} \cos \phi \cos\vartheta & 2^{-1/2}\sin \phi \\ 2^{-1/2}\sin \phi & \cos \phi \sin\vartheta \end{pmatrix}. \label{FormScatteringMatrix}
\end{equation}
Here, $\phi \in [0, \pi/2]$ describes the ratio of intra- to inter-band scattering and $\vartheta$ sets the ratio of the two intra-band scattering amplitudes. 
Using the general expression for $\zeta$ derived in Ref.~\cite{Timmons2020}, it is straightforward to compute $\zeta = \zeta(\vartheta,\phi)$ for the superconducting two-band model extracted above from the penetration depth measurements, see Methods. The result is shown in \figref{theoryfigure}(e). As opposed to $\lambda(T)$, the sensitivity parameter $\zeta$ is sensitive to the relative sign of the superconducting order parameter between the two bands. This sign  depends on the sign of $J$---a repulsive interband Cooper channel interaction $J$ leads to an $s^{+-}$ with opposite sign of the superconducting order parameter in the two bands while an attractive one leads to a $s^{++}$ state with the same sign. However, as a consequence of the strong momentum dependence of the absolute value of the order parameter on each band and between the two bands (captured by the anisotropy parameter $a$), the difference in the disorder sensitivities of $s^{++}$ and $s^{+-}$ is rather small. In fact, both scenarios can reproduce the rather large disorder sensitivity observed in experiment for natural values of $\phi$ close to but a bit smaller than $\pi/4$ [whereas $\vartheta$, related to the width of the red and blue regions in \figref{theoryfigure}(e) only plays a minor role]; these values of $\phi$ correspond to the generically expected scenario that local defects can induce both inter- and intra-band scattering processes with roughly equal magnitude. As such, the observed rather large disorder sensitivity, while surprising at first sight, further confirms the anisotropic two-band model extracted from the temperature-behavior of the penetration depth.

Going beyond the weak disorder limit, $\Gamma \rightarrow 0$, we see in \figref{fig3} that the $T_c$ reduction rate decreases with disorder strength. This is also consistent with a superconducting state transforming under a trivial representation but exhibiting significant ``accidental'' variation of its order parameter on and between different Fermi surfaces, such as the state we propose above. The observation of a significant but saturating change of $T_c$ upon increasing the amount of non-magnetic disorder combined with the aforementioned substantial increase of the $H_{c2}$ favor the anisotropic $s^{++}$ state over the $s^{+-}$ scenario.

In summary, our study of the temperature variation of the London penetration depth $\lambda(T)$ in the Dirac semimetal 2M-WS$_{2}$ reveals a low-temperature behavior $\lambda(T)-\lambda(0) \propto T^3$ down to $0.4$~K, clearly distinguishable from the exponential attenuation in a fully gapped isotropic $s$-wave BCS superconductor. In combination with the behavior close to the critical temperature $T_c$, this strongly indicates an anisotropic superconducting state with two weakly coupled gaps. In line with this conclusion, electron irradiation shows that the superconducting transition temperature of 2M-WS$_{2}$ decreases substantially upon induced non-magnetic disorder. Whereas these observations are consistent with both an $s^{++}$ and an $s^{+-}$ state, the behavior of the penetration depth and upper critical field with irradiation point towards an $s^{++}$ state. Our results motivate further work into the impact of such a highly anisotropic bulk superconducting state on proximity induced superconductivity of the topological surface state in 2M-WS$_2$.

\begin{acknowledgments}
We thank M.S.~Foster for discussions.
This work was supported by the U.S. Department of Energy (DOE), Office of Science, Basic Energy Sciences, Materials Science and Engineering Division. Ames Laboratory is operated for the U.S. DOE by Iowa State University under contract DE-AC02-07CH11358. Y. F. and F. H. were supported by the Shanghai Rising-Star Program (23QA1410700) and the National Natural Science Foundation of China (52103353). SYX and DB were supported by the NSF Career DMR-2143177. Work in France was supported by “Investissements d’Avenir” LabEx PALM (ANR-10-LABX-0039-PALM). The authors acknowledge support from the EMIR\&A French network (FR CNRS 3618) on the platform SIRIUS. P.P.O. acknowledges support from the
Research Corporation for Science Advancement via a Cottrell Scholar
Award. M.S.S.~acknowledges funding by the European Union (ERC-2021-STG, Project 101040651---SuperCorr). Views and opinions expressed are however those of the authors only and do not necessarily reflect those of the European Union or the European Research Council Executive Agency. Neither the European Union nor the granting authority can be held responsible for them. 
\end{acknowledgments}


%

\vspace{5em}

\begin{center}
\textbf{Methods}
\end{center}

\noindent\textbf{Penetration depth.} For a given temperature- and, in general, momentum- and band-dependent superconducting order parameter $\Delta_{j,\vec{k}}$, we compute the penetration depth $\lambda(T)$ from the expression \cite{EinzelSuperfluidDensity}
\begin{equation}
    \rho = 1 -  \sum_j \hat{D}_j \left\langle\int_{\frac{|\Delta_{j,\vec{k}}|}{T}}^\infty \textrm{d} x \frac{1}{2\cosh^2   \frac{x}{2}} \frac{x}{\sqrt{x^2 - \frac{|\Delta_{j,\vec{k}}|^2}{T^2}}}\right\rangle, \label{ExpressionForRho}
\end{equation}
for the normalized superfluid density $\rho(T) = (\lambda(0)/\lambda(T))^2$. Here $\braket{f_{\vec{k}}} \equiv \frac{1}{N_\Lambda}\sum_{\vec{k}\in\text{FS}} f_{\vec{k}}$ denotes an angular average over the direction of $\vec{k}$, normalized such that $\braket{1} = 1$, and $\hat{D}_j$ is the relative total density of states at the Fermi surface $j$ (obeying $\sum_j\hat{D}_j=1$). Here, we have assumed a homogeneous electromagnetic response tensor and have taken the Fermi velocity to exhibit a constant magnitude on each Fermi surface. For the fits of $\Delta\lambda(T)$ in \figref{theoryfigure}, we use $\Delta\lambda(T) = \lambda(0) [1/\sqrt{\rho(T)}-1]$ which yields the zero-temperature penetration depth $\lambda(0)$ as another fit parameter.

For the isotropic BCS model [$|\Delta_{\vec{k}}| = \Delta_0(T)$; only one band, thus, no $j$ index] and the anisotropic single-band model [$|\Delta_{\vec{k}}| = \Delta_0(T) (1 - a \cos 2\theta_{\vec{k}})$] of the main text, we use a BCS temperature dependence for $\Delta_0(T)$. For the two-band ($j=1,2$) model, we take $|\Delta_{j,\vec{k}}| = \Delta_j(T) (1 - a \cos 2 \theta_{\vec{k}})$, where $\Delta_j(T)$ are the solutions of the self-consistency equations [energies are measured in units of the cutoff scale, e.g., the Debye energy, which manifestly drops out in \equref{ExpressionForRho}]
\begin{equation}
    \Delta_j = \sum_{j'=1,2} \begin{pmatrix} U_1 & J_1 \\ J_2 & U_2 \end{pmatrix}_{j,j'} \Delta_{j'} \mathcal{I}(\Delta_{j'},T),
\end{equation}
where $U_{j}$ and $J_j$ are the intra- and inter-band Cooper-channel interaction constants (with positive referring to attractive interactions) multiplied by the respective density of states and
\begin{equation}
    \mathcal{I}(\Delta,T) = \int_0^1 \textrm{d} x \, \frac{\tanh \frac{\sqrt{x^2 + |\Delta|^2}}{2T}}{\sqrt{x^2 + |\Delta|^2}}
\end{equation}
is a dimensionless integral. To keep the number of fitting parameters at a minimum, we make the natural (consistent with $\hat{D}_1 \approx \hat{D}_2$) assumption $J_1 = J_2 \equiv J$ and $\hat{D}_j = U_j/(U_1 + U_2)$.

\vspace{1em}    
\noindent\textbf{Disorder sensitivity.} To compute the impact of weak disorder, we use the general expression of Ref.~\onlinecite{Timmons2020} for the disorder sensitivity $\zeta$, quantifying the disorder-induced change $\delta T_c$ of the critical temperature $T_c$ via 
\begin{equation}
    \delta T_c \sim -\frac{\pi}{4} \tau^{-1} \zeta 
\end{equation}
as the scattering rate $\tau^{-1} \rightarrow 0$: denoting the matrix elements of the impurities by $W_{\vec{k},\vec{k}'}$, which are $2 \times 2$ matrices in band space (we suppress spin) in our case, and defining the commutator (we here focus on non-magnetic impurities) $C_{\vec{k},\vec{k}'} = \mathcal{D}_{\vec{k}} W_{\vec{k},\vec{k}'}- W_{\vec{k},\vec{k}'} \mathcal{D}_{\vec{k}'}$, where $\mathcal{D}_{\vec{k}}$ is the superconducting order parameter (again a $2\times 2$ matrix in our case), it holds
\begin{equation}
    \zeta = \frac{\sum_{\vec{k},\vec{k}'\in\text{FS}} \text{tr}\left[ C^\dagger_{\vec{k},\vec{k}'} C^\pdagger_{\vec{k},\vec{k}'}  \right]}{2 \sum_{\vec{k},\vec{k}'\in\text{FS}} \text{tr}\left[ W^\dagger_{\vec{k},\vec{k}'} W^\pdagger_{\vec{k},\vec{k}'}\right] \sum_{\vec{k}\in\text{FS}} \text{tr}\left[\mathcal{D}_{\vec{k}}^\dagger \mathcal{D}_{\vec{k}}^\pdagger \right]}. \label{FormOfZeta}
\end{equation}
While the form of \equref{FormOfZeta} is independent of the basis, we will evaluate it in the band basis, where it holds $\mathcal{D}_{\vec{k}} = \chi_{\vec{k}} \text{diag}(\Delta_1,\Delta_2)$ with $\chi_{\vec{k}} = 1 - a \cos 2 \theta_{\vec{k}}$ in our model. Restricting the discussion to local impurities with $W_{\vec{k},\vec{k}'} = w$ as given in \equref{FormScatteringMatrix}, it is straighforward to evaluate \equref{FormOfZeta} analytically; the result is plotted in \figref{theoryfigure}(e) for the two-band model.


\end{document}